\documentclass[12pt,a4paper]{article}
\usepackage{amsmath}
\usepackage{amsfonts}
\usepackage{amssymb}
\usepackage{graphics}
\usepackage{epsfig}
\usepackage{color,amsmath,bm}
\usepackage{graphicx}
\usepackage{subfigure}

\newcommand{\beq}{\begin{equation}}
\newcommand{\eeq}{\end{equation}}

\begin{document}

\begin{titlepage}
\renewcommand{\thefootnote}{\fnsymbol{footnote}}
\begin{flushright}
{\tt FTPI-MINN-07/32; UMN-TH-2623/07;\\
 IFUP-TH/2007-32;}
\end{flushright}
\begin{center}
{\Large  {\bf A Note on Chern-Simons Solitons}\\[10pt]
\large {\bf - a type III vortex from the wall vortex -} 
}

\renewcommand{\thefootnote}{\fnsymbol{footnote}}
\bigskip
{\large\bf
S.\
Bolognesi$^{\,\diamond}$\footnote[0]{$^{\diamond\,}$bolognesi(at)physics.umn.edu}
{\small and} S.~B.~Gudnason$^{\,\natural}$\footnote[0]{$^{\natural\,}$gudnason(at)df.unipi.it} \\[2mm]
{\small \sl ${\diamond\,}$William I. Fine Theoretical Physics Institute,
University of Minnesota,}\\[-1mm]
{\small \sl  116 Church St. S.E., Minneapolis, MN 55455, USA}\\
{\small \sl $^{\natural\,}$ Dipartimento di Fisica, ``E.~Fermi'',
  University of Pisa, and}\\[-1mm]
{\small \sl Istituto Nazionale di Fisica Nucleare, Sezione di Pisa,}\\
{\small \sl Largo Pontecorvo, 3, Ed.~C, 56127 Pisa, Italy} }
\end{center}

\renewcommand{\thefootnote}{\arabic{footnote}}
\setcounter{footnote}{0}

\bigskip
\noindent
\begin{center} {\bf Abstract} \end{center}

We study some properties of topological Chern-Simons vortices in $2+1$
dimensions. As has already been understood in the past, in the large
magnetic flux limit, they are well described by a Chern-Simons domain
wall, which has been compactified on a circle with the symmetric phase
inside and the asymmetric phase on the outside.  

Our goal is two-fold. First we want to explore how the tension depends
on the magnetic flux discretized by the integer $n$. The BPS case is 
already known, but not much has been explored about the non-BPS
potentials.  
A generic renormalizable potential has two dimensionless parameters that can be
varied.  Variation of only one of them lead to a type I and type II
vortex, very similar to the Abrikosov-Nielsen-Olesen (ANO)
case. Variation of both the parameters leads to a much richer
structure. In particular we have found a new type of vortex, which is
type I-like for small flux and then turns type II-like for larger
flux. We could tentatively denote it a type III vortex. 
This results in a stable vortex with number of fluxes which can be
greater than one.

Our second objective is to study the Maxwell-Chern-Simons theory and
and understand how the large $n$ limit of the CS vortex is smoothly
connected with the large $n$ limit of the ANO vortex.

\bigskip
\vfill
\begin{flushleft}
June, 2008
\end{flushleft}
\end{titlepage}

\section{Introduction}

Physics in $2+1$ dimensions has many interesting aspects. In
particular, there is the possibility of particles carrying fractional
spin and thus behaving as anyons \cite{librowilc}. Strictly related to
the latter property it is possible, in $2+1$ dimensions, to introduce
a topological Chern-Simons term in the Lagrangian. The Gauss law in
the presence of this term reads
\begin{equation}
\frac{\kappa}{2}\epsilon^{\mu\nu\rho}F_{\nu\rho}=eJ^{\mu} \ ,
\label{gausslaw}
\end{equation}
where $\kappa$ is the coefficient of the Chern-Simons term. This, in
particular, implies $B=F_{12}=\frac{eJ^0}{\kappa}$ : a magnetic flux is
permanently attached
to a charged particle. Through the Ahranov-Bohm effect this,
effectively, attaches an extra spin to charged particles. This
statistical transmutation is a crucial property in the effective
explanation of the fractional quantum Hall effect (FQHE). Electrons
can effectively be described by bosons plus an opportune
Chern-Simons interaction. Condensation leads to a superconducting
quantum Hall fluid \cite{librpFQHE,KHKmodel}.

Solitons, and in particular charged vortices, play an essential role
in $2+1$ dimensional physics and acquire interesting properties when a
Chern-Simons term is present in the Lagrangian. First of all, being
magnetic vortices, they also acquire an electric charge due to the
Chern-Simons term. This makes them be anyons, and their semi-classical
nature makes them suitable candidates to the study of anyon dynamics. They
also appear as essential ingredients in the effective descriptions
of the FQHE. In particular, when the magnetic field is increased
between two levels of the quantum Hall fluid, vortices are created
to compensate the extra magnetic field.

The literature concerning Chern-Simons solitons is quite rich.
The Chern-Simons term, while not being dynamical, acquires physically
different
properties when coupled to different kinds of matter fields. A first
obvious distinction is if the matter is relativistic or
non-relativistic. The latter is certainly more appropriate when
considering condensed matter phenomena. Another distinction is if
the gauge group is Abelian or non-Abelian. Finally, there is also the
possibility of having (or not having) the Maxwell (or Yang-Mills) term
to supply kinetic properties to the gauge field. In this paper, we
shall restrict ourselves to an Abelian gauge field coupled to
relativistic matter, with and without the Maxwell term.

In recent works \cite{miei,conbjarke,monopolo}, we have considered
the behavior of topological solitons, vortices and monopoles, when
the magnetic flux becomes very large. Although not a priori
expected, there is a simple realization of this large $n$ limit of
solitons. A domain wall appears as a sub-structure, separating an
internal phase, an unstable point of the potential where the
symmetry is restored, from the external vacuum. Stabilization is
achieved by a balance of energies between the magnetic flux, where
the force is repulsive, and the scalar fields, for which it is
attractive.

The purpose of this project is to study the large flux limit of
Chern-Simons solitons and see how this is related to the previous
ones. The simplest example of a self-dual Chern-Simons vortex is
obtained in the so-called Chern-Simons-Higgs system. It consists of
an Abelian Chern-Simons term coupled to a scalar field with a
suitable sixth-order potential \cite{Jackiwprl}. The potential has
two degenerate vacua, one where the gauge symmetry is unbroken and
one where the Higgs field acquires an expectation value. This is a
basic and recurrent difference between Chern-Simons solitons and
ordinary ones. In particular, there exists a domain wall separating
the two vacua. The theory also admits solitons of various kinds in
each vacuum, both topological and non-topological. In the large flux
limit, vortices approach domain walls compactified on a circle. This
has first been observed in \cite{Jackiwesteso}. 

One problem we shall consider is that of non-BPS potentials. We
shall see that for particular values of the potential parameters,
interesting new properties appear. In particular, there is the
possibility of having a configuration
where the vortex which minimizes the tension per unit flux, for
certain parameters, can have an arbitrary number of fluxes and thus
an arbitrary size. This is interesting, because the number of stable
vortices is finite and the effective size of multiwinding vortices (or
coincident vortices) hence is finite, but bigger than the classic
stable 1-vortex of a (normal) type II superconductor.

Having in mind the study of the relation between the large $n$ limit of the
Maxwell vortex \cite{conbjarke} and the Chern-Simons one
\cite{Jackiwesteso}, we proceed to consider the Maxwell-Chern-Simons
model \cite{Maxwellchernsimons}. This self-dual model contains both
the Maxwell
and the Chern-Simons term for an Abelian gauge field. It is coupled
to two scalar fields, one real and one complex. This model is
particularly rich and it can, by changing the values of the
parameters, interpolate between the pure Maxwell and the pure
Chern-Simons
theories. It is thus the right place to study the relation between
the two large $n$ limits. Even in this model, there are two
degenerate vacua, one symmetric and one Higgsed. We expect that a
large Maxwell-Chern-Simons vortex becomes a domain wall compactified
on a circle. The matching of the large $n$ limits of pure Maxwell and
pure Chern-Simons theories will be discussed.

We have divided the paper into two sections. In the next section we
first give a review of the Chern-Simons-Higgs model and then we will
study the construction of the wall-vortex, that is, the large $n$
description of the vortex in this system. 
In Section \ref{secondasezione}, we take into account the
generalization with the Maxwell term and study again the (expected)
large $n$ substructure of the vortex. 
We conclude in Section \ref{conclusione} and comment about 
possible future developments. During this project we have made
extensive use of the review \cite{Dunnereview}.

\section{Abelian Chern-Simons-Higgs System}

\label{primasezione}

We will first make a short review of the Abelian Chern-Simons-Higgs
system \cite{Jackiwprl,Jackiwesteso} and it is obtained by combining
the Abelian Chern-Simons term and a scalar Higgs field 
\begin{equation}
\mathcal{L}=\frac{1}{4}\kappa\epsilon^{\mu\nu\rho}A_{\mu}F_{\nu\rho}+\left%
\vert D_{\mu}\phi\right\vert ^{2}-V\left(\left\vert \phi\right\vert \right)
\ ,   \label{start}
\end{equation}
where the covariant derivative is $D_{\mu}\phi=\left(
\partial_{\mu}-ieA_{\mu}\right) \phi$, the space-time metric is 
$\eta_{\mu\nu}=\mathrm{diag}\left( 1,-1,-1\right)$.
The self dual potential is
\begin{equation}
V\left(\left\vert \phi\right\vert \right) =\frac{e^4}{\kappa^{2}}\left(
\left\vert \phi\right\vert ^{2}-v^{2}\right) ^{2} \left\vert \phi\right\vert
^{2} \ . \label{self-dual-potential}
\end{equation}
The theory possesses two vacua : $\phi = 0$ and $|\phi| = v$, one
symmetric and one asymmetric.

The energy of the system is
\begin{align}
E & =\int d^{2}x\left[ \left\vert D_{0}\phi\right\vert ^{2}+\left\vert
D_{i}\phi\right\vert ^{2}+V\left( |\phi|\right) \right]\ , \nonumber\\
& =\int d^{2}x\left[ \left\vert \partial_{0}\phi\right\vert ^{2}+\frac {%
\kappa^{2}B^{2}}{4e^2|\phi|^{2}}+\left\vert D_{i}\phi\right\vert ^{2}+V\left(
|\phi|\right) \right] \ .
\end{align}
The term $\frac{\kappa^{2}B^{2}}{4e^2|\phi|^{2}}$ is fundamental, it
forces the 
magnetic field to stay away from the $\phi=0$ region, that is, the
magnetic field has to go to zero faster than $|\phi|^2$. Hence, the
magnetic flux will be concentrated on the boundary between the two
vacua (as it cannot sustain in the Higgs vacuum).

The equations of motion are
\begin{align}
D_{\mu}D^{\mu}\phi & =-\frac{\delta V}{\delta\phi^{\ast}}~,
\label{eqomsecondorder} \\
\frac{1}{2}\kappa\epsilon^{\mu\nu\rho}F_{\nu\rho} & = -ie\left(
\phi^{\ast}D^{\mu }\phi-\phi D^{\mu}\phi^{\ast}\right) \
. \label{gausslaw2}
\end{align}
In particular, the last equation is the Gauss law
(\ref{gausslaw}). The electromagnetic field is non-dynamical since
there is no Maxwell term and the time component of the gauge field is
thus algebraically determined
\beq A_0 = -\frac{\kappa}{2e^2}\frac{F_{12}}{|\phi|^2} \
. \label{gausslaw3}\eeq
Hence, it will enter the Hamiltonian only through the covariant
derivative.

A comment on the phases and the perturbative spectrum in the
two vacua of the theory :  when $\phi=0$, the scalar field describes a
charged particle with mass $\mu=\frac{e^2v^2}{\kappa}$. The Chern-Simons
term has the effect of changing the spin of these particles 
\cite{Jackiwprl,Jackiwesteso,Banerjee:1996we}. Since
they have charge $1$, the Gauss law implies that they have
magnetic flux $\frac{e}{\kappa}$. The Aharonov-Bohm phase gives them
an ``effective'' spin equal to $\frac{e^2}{4\pi\kappa}$. In the Higgs
phase $\phi=v$, there is a neutral scalar particle of mass
$m\equiv\frac{2e^2v^2}{\kappa}$. The gauge field is massive and thus there is
no effect of spin transmutation.

Performing the Bogomol'nyi factorization \cite{Bogomolny:1975de} one obtains%
\begin{align}
E & =\int d^{2}x\left[ \left\vert D_{0}\phi\pm\frac{ie^2}{\kappa}\left( |\phi
|^{2}-v^{2}\right) \phi\right\vert ^{2}+\left\vert D_{\pm}\phi\right\vert
^{2} \pm ev^{2}B\right] \ , \label{CSBPS}
\end{align}
where $D_{\pm}\phi \equiv \left(D_1\pm iD_2\right)\phi$. 
Thus, there exists the following bound on the energy
\begin{equation}
E\geq ev^{2}|\Phi_{B}|  \ , \label{BPSbound}
\end{equation}
where $\Phi_B \equiv \int d^2x\, B = 2\pi n/e$ is the magnetic flux and
$n$ denotes the vortex number.

We will now describe what we are going to do. First we will review
the domain wall solution that interpolates between the two
vacua and subsequently the domain wall with the addition of a magnetic
flux. Then in Section \ref{uno}, we will show how the Chern-Simons
vortices behave as compactified walls in the large flux limit. We
shall then consider, in detail, the topological vortex and provide
numerical evidence for this claim in Section \ref{due}. Finally, we
consider the behavior of topological vortices when the potential is
not the BPS one in Section \ref{quattro}.

\subsection{The domain wall solution and the large flux
  limit\label{Sec-CSDW}}\label{uno}

Now we will review the domain wall in the Chern-Simons model, first
without flux and subsequently with added flux \cite{Jackiwesteso,Kao:1996tv}.
The Bogomol'nyi factorization for the domain wall is%
\begin{align}
T & =\int dx\left[ |\partial_{x}\phi|^{2}+\frac{e^4}{\kappa^{2}}\left( \left\vert
\phi\right\vert ^{2}-v^{2}\right) ^{2}\left\vert \phi\right\vert ^{2}\right]
\ , \nonumber\\
& =\int dx\left[ \left\vert \partial_{x}\phi\pm\frac{e^2}{\kappa}\left(
  \left\vert
\phi\right\vert ^{2}-v^{2}\right) \phi\right\vert ^{2} \mp \frac{e^2}{2\kappa}
\partial_x (\left\vert\phi\right\vert^2-v^2)^2 \right] \ , \label{basiccswall}
\end{align}
hence, the wall tension and the wall configuration which
saturates the BPS bound read
\begin{equation}
T_{\text{\textrm{wall}}}=\frac{e^2v^{4}}{2\kappa}~,\qquad\phi_{\mathrm{wall}}\left(
x\right) =\frac{v}{\sqrt{1+e^{-m\left( x-x_{0}\right) }}} \ ,
\label{wall profile}
\end{equation}
where $m \equiv \frac{2e^2v^2}{\kappa}$.

One can add a magnetic flux to the domain wall, by switching on a
gauge field $A_y(x)$ such that $A_y(-\infty)=-f$ and
$A_y(+\infty)=0$. The result is a magnetic flux density equal to $f$
\cite{Jackiwesteso}. The wall tension 
can be rewritten as 
\begin{align}
T =\int dx\bigg[ &\left\vert \partial_{x}\phi\pm\frac{e^2}{\kappa}\left(
\left\vert \phi\right\vert ^{2}-v^{2}\right) \phi\right\vert^{2} +
\left\vert eA_y
\phi \pm' \frac{\kappa}{2e}\frac{\partial_xA_y}{\phi^*}\right\vert^2 
\notag \\
& \mp
\frac{e^2}{%
2\kappa} \partial_x\left(\left\vert\phi\right\vert^2-v^2\right)^2
 \mp' \frac{\kappa}{2}\partial_x\left(A_y^2\right)%
\bigg] \ .
\end{align}
The gauge field does not change the equation for the scalar field
which thus remains unchanged. The second set of $\pm$s are independent
of the first set and are thus marked with a prime. The magnetic field
gives a contribution to the wall tension which now becomes\footnote{
This meanwhile introduces a momentum density in the $y$-direction
equal to $\frac{\kappa f^2}{2}$.}
\begin{equation}  \label{tensione}
T_{\rm wall+flux} =\frac{e^2v^4}{2\kappa}+\frac{\kappa f^2}{2} \ .
\end{equation}

\begin{figure}[h!tb]
\epsfxsize=7cm \centerline{\epsfbox{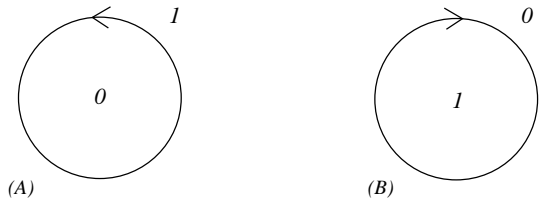}}
\caption{{\protect\small Schematic representation of the basic
spherical symmetric solitons in the Chern-Simons-Higgs theory. They are
both made of the domain wall compactified on a circle and stabilized
by the angular momentum. We denote the symmetric vacuum by $0$ and the
asymmetric vacuum by $1$. According to whether the vacuum $0$ (unbroken
phase) is inside or outside of the circle, we have respectively the
topological vortex or the non-topological soliton of
Ref.~\cite{Jackiwesteso}. In both cases, we 
have chosen the orientation of the magnetic flux out of the plane
(towards the reader).}} \label{prima}
\end{figure}

Having reviewed the domain wall with magnetic flux, we can now
construct the Chern-Simons wall vortex as follows. 
We consider the compactification of the wall with flux on a circle of
radius $R$ (i.e. along the $y$-direction). The
stabilization is achieved through a balance between the tension of the
wall and the energy due to the magnetic field. We have to keep in mind
that while varying the radius $R$, what remains constant is the total
magnetic flux $\Phi_B=2\pi R f$. The energy as function of the flux is
thus
\begin{equation}
\label{tensione1}
E(R)= \frac{e^2v^4 \pi R}{\kappa} + \frac{\kappa\Phi_B^2}{4\pi R} \ ,
\end{equation}
and the minimization of this system gives
\begin{equation}
R=\frac{\kappa\Phi_B}{2\pi ev^2} = \frac{e\Phi_B}{\pi m} \ , \qquad
E=ev^2\Phi_B \ . \label{largendwradius}
\end{equation}
This solution saturates the BPS bound (\ref{BPSbound}) of the vortex
system. This implies
that in the large flux limit of the vortex, the solution should
exactly become a compactified wall.
A useful remark in store, is that the flux density $f$ of the wall
vortex (the compactified wall, see Fig.~\ref{prima}) is simply
proportional to the mass: 
\footnote{
A remark on the topological vortex and the non-topological
soliton of Ref.~\cite{Jackiwesteso}. The only difference is in the
sign of the angular momentum which is obvious from Figure
\ref{prima}. The angular momentum can easily be computed to be $J=2\pi
R^2 P$ with $P$ the momentum density, 
which yields $J=\pm \frac{\kappa}{4\pi}\Phi_B^2$.}
\beq f = \frac{n}{eR} = \frac{m}{2e} \ . \label{flux-density} \eeq

\subsection{The topological vortex in the large flux limit}

\label{due}

Considering now the topological vortex in a cylindrical symmetric
ansatz \cite{Jackiwprl,Jackiwesteso}
\begin{align}
\phi =ve^{in\theta}h\left( r\right)\ ,   \qquad
eA_{\theta} = \frac{n}{r}a(r)\ . \label{cylindrical ansatz}
\end{align}
What we want to show is that the scalar field of the vortex solution
for large $n$, simply has the profile of the domain wall (\ref{wall
  profile}).

The profile functions $h\left( r\right) $ and $a(r)$ satisfy the following
boundary conditions: they both vanishes at $r=0$ and they both saturate at $1
$ for $r\rightarrow\infty$. After insertion of the ansatz into the BPS
equations of motion (\ref{CSBPS}) we get
\begin{align}
\frac{\partial_ra}{r} \pm \frac{m^2}{2n}\left(h^2-1\right)h^2 &= 0 \ ,
\label{cyldif1} \\
\partial_rh \pm \frac{n}{r}\left(a-1\right)h &= 0 \ ,  \label{cyldif2}
\end{align}
with $m=\frac{2e^2v^2}{\kappa}$ as defined previously. We are looking for two boundary conditions in
order to solve the problem numerically. The limiting behavior of the
functions as $r\to 0$ is
\beq
h = Ar^n \ , \qquad a = \frac{m^2}{2n(2n+2)}A^2r^{2n+2} \ , \label{leftbc}
\eeq
and in the limit $r\to\infty$
\beq
h = 1 - Fe^{-mr} \ , \qquad a = 1 - \frac{rmF}{n}e^{-mr} \ ,  \label{rightbc}
\eeq
where $A,F$ are constants.

From these equations, we can form the following boundary conditions
\begin{align}
\lim_{r\to 0} \left(a - \frac{m^2r^2}{2n(2n+2)}h^2\right) = 0 \ , \qquad
\lim_{r\to\infty} \left(a - 1 + \frac{rm}{n}(1-h)\right) = 0 \ .
\end{align}

\begin{figure}[!ht]
\begin{center}
\includegraphics[width=0.90\linewidth]{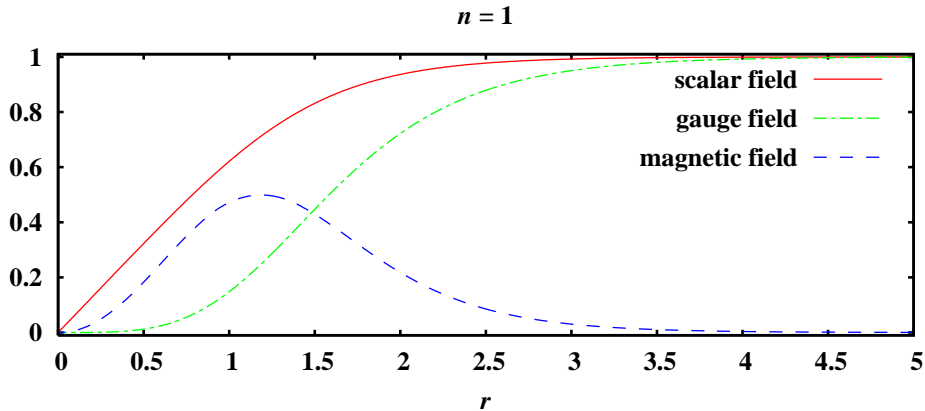}
\end{center}
\caption{{\protect\small Profile functions for the Chern-Simons vortex with $%
n=1$, $m=2$, where the red line (solid) is the scalar field profile,
the green line (dash-dotted) is the electromagnetic potential profile
(i.e. $a(r)$ given by (\ref{cylindrical ansatz}))
and the blue line (dashed) is the magnetic field. Notice that the
magnetic field is already for $n=1$ pushed completely outside $r=0$
and is thus always a ring of flux placed at the vortex boundary. }}
\label{sol2}
\end{figure}

For the numerical results we shall set the parameter $m = 2$, which
corresponds to the choice $e=\kappa=v=1$. The solution
for winding number $n = 1$ is shown in Figure \ref{sol2}.

\begin{figure}[h!tb]
\begin{center}
\includegraphics[width=0.90\linewidth]{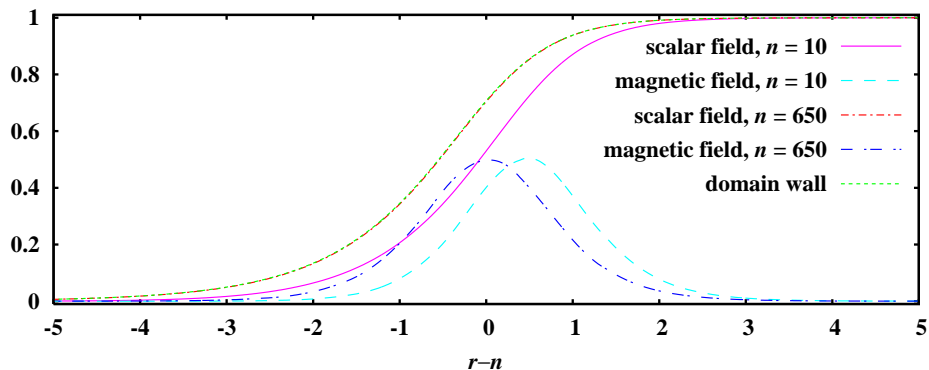}
\end{center}
\caption{{\protect\small Profile functions for Chern-Simons vortex for
various $n$. The radius of the vortex is $R_{\mathrm{vortex}} = n$ for
$m=2$. Notice that already for $n=650$, the scalar field and the
domain wall coincide. }}
\label{intermediate}
\end{figure}
\begin{figure}[h!tb]
\begin{center}
\includegraphics[width=0.90\linewidth]{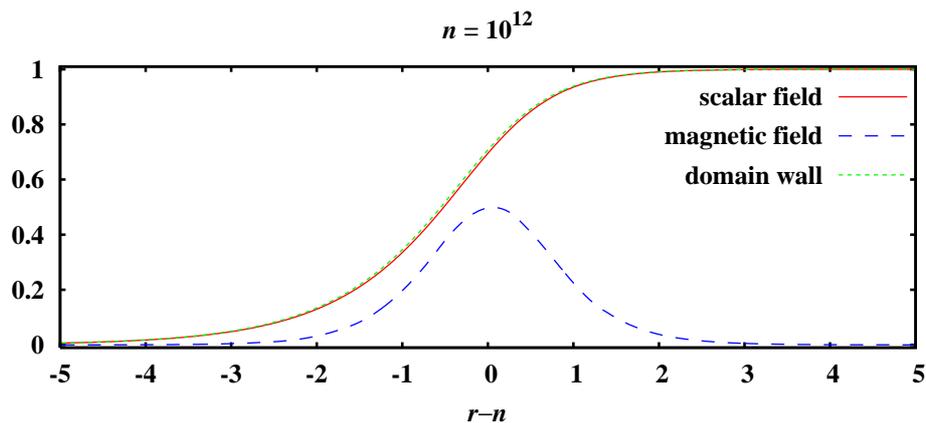}
\end{center}
\caption{{\protect\small Profile functions for the Chern-Simons vortex for $%
n=10^{12}$. Notice that the scalar field and the domain wall
coincide. }}
\label{bps10hat11}
\end{figure}

We then solve the equations for various values of $n$. The corresponding
profile functions are shown in Figure \ref{intermediate},
\ref{bps10hat11}.
It is observed that the profile functions of the vortex at large $n$
for the topological vortex are simply described 
by the profile function of the domain wall (\ref{wall profile})
situated at radius (\ref{largendwradius}). The magnetic field is then
simply obtained from (\ref{cyldif1}).

\subsection{Non-BPS potentials}

\label{quattro}

A generic renormalizable potential in $2+1$ dimensions is of sixth
order, and can be parameterized by two couplings $\alpha$ and $\beta$
\begin{equation}
V\left( \left\vert \phi\right\vert \right) =\frac{\alpha e^4}{\kappa^{2}}\left(
\left\vert \phi\right\vert ^{2}-v^{2}\right) ^{2}\left[ \left\vert
\phi\right\vert ^{2}-\beta\left( \left\vert \phi\right\vert
^{2}-v^{2}\right) \right] \ .
\end{equation}
When $\alpha=1$ and $\beta=0$ it corresponds to the BPS potential
(\ref{self-dual-potential}). We now want to consider the Chern-Simons
solitons in the case of a generic potential. We shall here concentrate
on the topological vortex.

Before doing the computation we will make some qualitative remarks in order
to understand what we should expect. Consider first the case $\beta =0$. In
this case the potential still has two degenerate vacua, we are only changing
the height of the potential. The large flux limit is very useful in order to
understand the qualitative behavior. In the large flux limit we still
expect the vortex to become a compactified wall. The analysis of
Section \ref{Sec-CSDW}
should thus go through unchanged, apart from a factor of
$\sqrt{\alpha}$ in front of the wall tension.

Keeping $\beta=0$ and varying only $\alpha$ should thus gives a phenomenology
very similar to that of the Abrikosov-Nielsen-Olesen (ANO) vortex
\cite{ano}. In 
the large flux limit the tension
is always asymptotically linear in $n$. Only for $\alpha=1$ this linear
dependence is exact for all values of $n$. In the intermediate regime the
vortices are of type I (attractive) or type II (repulsive) depending on
$\alpha$ being smaller or greater than one, respectively.

We now want to consider the case of $\beta$ different from zero. Since we
are focusing on the topological vortex, we choose $\beta<1$ so that the
Higgs phase remains the true vacuum while the Chern-Simons phase acquires a
vacuum energy density $\varepsilon_0 = \frac{\alpha \beta
 e^4 v^6}{\kappa^2}$ and is
metastable for $\beta < 1/3$ and an unstable extremum otherwise.

The energy function for the compactified wall now reads 
\begin{equation}
E(R)= T_{\mathrm{W}}(\alpha,\beta)2\pi R + \frac{\kappa\Phi_B^2}{4\pi
  R} +\varepsilon_0 \pi R^2 \ ,  \label{nonBPSenergy}
\end{equation}
where $T_{\mathrm{W}}(\alpha,\beta)$ is the domain wall tension as function of the parameter $\alpha$, $\beta$.
Now to obtain the wall vortex, we have to perform a minimization with
respect to the radius $R$, in the large $n$ limit. We will see in a
minute, that $R$ is large in the large $n$ limit, thus we can neglect
the first term in (\ref{nonBPSenergy}) and the result is
\beq R = \sqrt[3]{\frac{\kappa\Phi_B^2}{8\pi^2 \varepsilon_0}}\ ,
\qquad
E = \frac{3}{4}\sqrt[3]{\frac{\kappa^2\varepsilon_0\Phi_B^4}{\pi}} \
, \label{nonBPSlargefluxformulae}\eeq
which is equivalent to the vortex radius and energy in the the large
flux limit and furthermore, in terms of $n$, the radius and energy
read $R \propto n^{2/3}$ and $E\propto n^{4/3}$. Consistently, the
radius goes as $n^{2/3}$, (i.e.~the first term in (\ref{nonBPSenergy})
goes like $R$ while the two remaining terms go like $R^2$) and thus
our previous approximation holds in the large $n$ limit.  

The equations of motion for the non-BPS Abelian Chern-Simons vortex are
\begin{align}
\partial_r^2a - \frac{1}{r}\partial_ra
-\frac{2\left(\partial_ra\right)\left(\partial_rh\right)}{h}
+m^2\left(1-a\right)h^4 &= 0 \ , \\
\frac{1}{r}\partial_r\left(r\partial_rh\right) - \frac{n^2}{r^2}%
\left(1-a\right)^2h +
\frac{n^2\left(\partial_ra\right)^2}{m^2r^2h^3} -
\frac{1}{2v^2}\frac{\partial V}{\partial h} &= 0 \ .
\end{align}
For a generic potential we obtain
\begin{equation}
\frac{1}{2v^2}\frac{\partial V}{\partial h} = \frac{\alpha}{4}%
m^2\left(h^2-1\right)\left[3h^2- 3\beta\left(h^2-1\right)-1\right]h
\ .
\end{equation}
In order to find numerical solutions, we need the boundary conditions at $%
r\to 0$ and $r\to\infty$. The limiting behaviors of the profile
functions are for $r\to 0$
\begin{equation}
h = Ar^n \ , \qquad a = Br^{2n+2} \ ,
\end{equation}
and for $r\to\infty$
\begin{equation}
h = 1-Fe^{-\sqrt{\alpha}mr} \ , \qquad a = 1-Ge^{-mr} \ .
\end{equation}
From these behaviors we can form the following conditions
\begin{align}
\lim_{r\to 0}\left(nh-rh^{\prime }\right) &= 0 \ , & \qquad
\lim_{r\to
0}\left((2n+2)a-ra^{\prime }\right) &= 0 \ , \\
\lim_{r\to\infty}\left(h + \frac{h^{\prime }}{\sqrt{\alpha}m}\right)
&= 1 \ , & \qquad \lim_{r\to\infty}\left(a + \frac{a^{\prime
}}{m}\right) &= 1 \ .
\end{align}
The vortex tension reads\footnote{Notice that in our abuse of
  notation, the tension is a 1-dimensional integral for the wall and
  2-dimensional integral for the vortex.}
\begin{align}
T = 2\pi v^2\int dr\,r\bigg\{ &\frac{n^2}{m^2}\frac{%
\left(\partial_ra\right)^2}{r^2h^2} + \left(\partial_rh\right)^2 + \frac{n^2%
}{r^2}\left(1-a\right)^2h^2 \notag \\
& + \frac{\alpha}{4}m^2\left(h^2-1\right)^2 \left[h^2-\beta\left(h^2-1\right)%
\right]\bigg\} \ .
\end{align}

First, we study numerically the system with $\beta=0$. The energy
function of the compactified wall will, however, change with respect to
(\ref{nonBPSenergy}) and (\ref{nonBPSlargefluxformulae}) because of
zero $\varepsilon_0$ and it is
\beq E(R) = \frac{\sqrt{\alpha}e^2v^4\pi R}{\kappa} +
\frac{\kappa\Phi_B^2}{4\pi R}\ , \eeq
which gives a radius and energy in the large flux limit of 
respectively
\beq R = \frac{\kappa\Phi_B}{\sqrt[4]{\alpha}2\pi ev^2} \ , \qquad
E = \sqrt[4]{\alpha}ev^2\Phi_B \ . \eeq
In the large flux limit, we can calculate the profile function for the
scalar field and the gauge field analytically, using the $1+1$
dimensional system (the domain wall)
\begin{align}
\lim_{n\to\infty} h =
\frac{1}{\sqrt{1+e^{-\sqrt{\alpha}m(x-x_0)}}} \ , \qquad
\lim_{n\to\infty} a =
\frac{e^{m(x-x_0)}}{\left[1+e^{\sqrt{\alpha}m(x-x_0)}\right]^{\frac{1}{\sqrt{\alpha}}}}
\ .
\end{align}
We find exact agreement of the numerical integrated profile functions
for large values of $n$ with the above results; the vortex becomes a
compactified wall in the large $n$ limit.

\begin{figure}[h!tb]
\begin{center}
\includegraphics[width=0.68\linewidth]{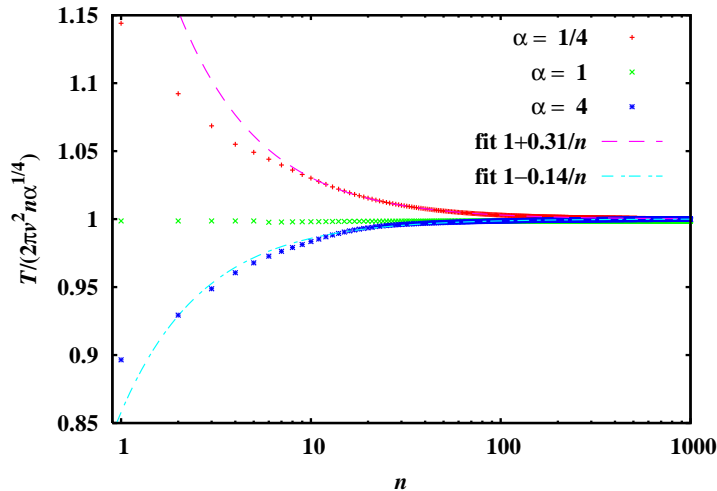}
\end{center}
\caption{{\protect\small Vortex tension divided by $2\pi v^2
    \sqrt[4]{\alpha}\,n$ as function of the winding number $n$, for
    various values of $\alpha$; $1/4, 1, 4$, corresponding to a type
    I, a BPS and a type II vortex. $1/n$ fits of the non-BPS vortex
    tensions are shown and are seen to match reasonably at large $n$. }}
\label{cstensionbetazero}
\end{figure}

In Figure \ref{cstensionbetazero} is shown the vortex tension
normalized for convenience by a numerical factor and
$\sqrt[4]{\alpha}$, which puts the different vortex tensions on equal
footing at large $n$. 
We denote by $\mathcal{T}\equiv\frac{T}{n}$ the vortex tension per
unit flux. The curves for the vortex tension per unit flux
$\mathcal{T}$ approach the large flux limit value (i.e. $2\pi
v^2\sqrt[4]{\alpha}$) approximately as $1/n$ (see Figure
\ref{cstensionbetazero}). 

We will now turn to the generic case with $\beta\neq 0$. In terms of $n$,
the vortex tension per unit flux $\mathcal{T}$ will go as $n^{\frac{1}{3}}$
in the large flux limit. 
First a word of our expectations. We seek to combine the type I vortex
behavior at small $n$ (attractive force) with the large $n$ behavior
due to the presence of a non-zero vacuum energy density $\varepsilon_0
= \frac{\alpha\beta e^4v^6}{\kappa^2}$. Naively, this gives an attractive
force for small $n$ and a repulsive force for large $n$ and thus a
vortex with finite units of flux greater than one (finite size) as the
ones with additional flux will decay.

\begin{figure}[h!tb]
\begin{center}
\includegraphics[width=0.68\linewidth]{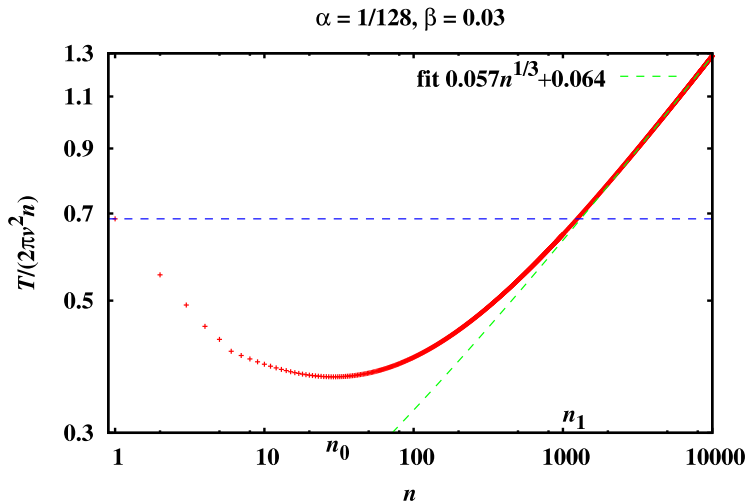}
\end{center}
\caption{{\protect\small Vortex tension divided by $2\pi v^2 n$ as function
    of the winding number $n$, for a vortex with $\alpha =
    \frac{1}{128}$ and $\beta = 0.03$. The large $n$ behavior
    is as predicted proportional to $n^{\frac{1}{3}}$ and the small
    $n$ behavior is type I-like, thus we have found a vortex with
    attractive force for small $n$ and repulsive force for large
    $n$. We could tentatively denote it a type III vortex. $n_0$
    denotes the winding number with the minimal vortex tension. }}
\label{nonzerobetaprima}
\end{figure}

In Figure \ref{nonzerobetaprima}, we show the vortex tension per unit
of flux for a type I 
($\alpha < 1$) vortex with $\alpha=\frac{1}{128}$
where we switch on a small $\beta = 0.03$. For this value of $\beta$,
the Chern-Simons vacuum is metastable.
\bigskip

From the figure we can define three domains
\begin{align}
A &: 1 < n \le n_0 \ , \ {\rm where} \ T(n_0) < T(n) \ , \forall n\neq
n_0 \ , \nonumber\\
B &: n_0 < n < n_1 \ , \ {\rm where} \ n_1 \equiv \left\{n'\in\mathbb{Z}_+\,|\,
  \min\left(\frac{T(n')}{n'}\right) \ge T(1) \right\} \ , \\
C &: n_1 \le n \ , \nonumber
\end{align}
where we have assumed no degeneracy of the lowest tension
state. Considering first the domain $A$, we can prove stability as
follows $T(2)<2T(1),$ is stable; $T(3)<3T(1),$ and
$\frac{T(3)}{3}<\frac{T(2)}{2} \Rightarrow T(3)<
T(2)+\frac{T(2)}{2}<T(2)+T(1),$ thus it is stable in all channels.

Generically
\begin{align}
T(n+m) &< T(n) + \frac{m}{n}T(n) \ , \ {\textrm{for}} \ n+m \le n_0 \ ,
\nonumber\\
&< T(n) + mT(1) \ , \\
&< T(n) + T(m) \ , \quad {\rm for} \ , \ m \le n \ ,  \nonumber
\end{align}
where $n,m \in \mathbb{Z}_+$.
Hence, by induction it is seen that the vortices in domain $A$ are
stable to decay in any channel. 
\begin{figure}[h!tb]
\begin{center}
\mbox{\subfigure
{\includegraphics[width=0.7\linewidth]{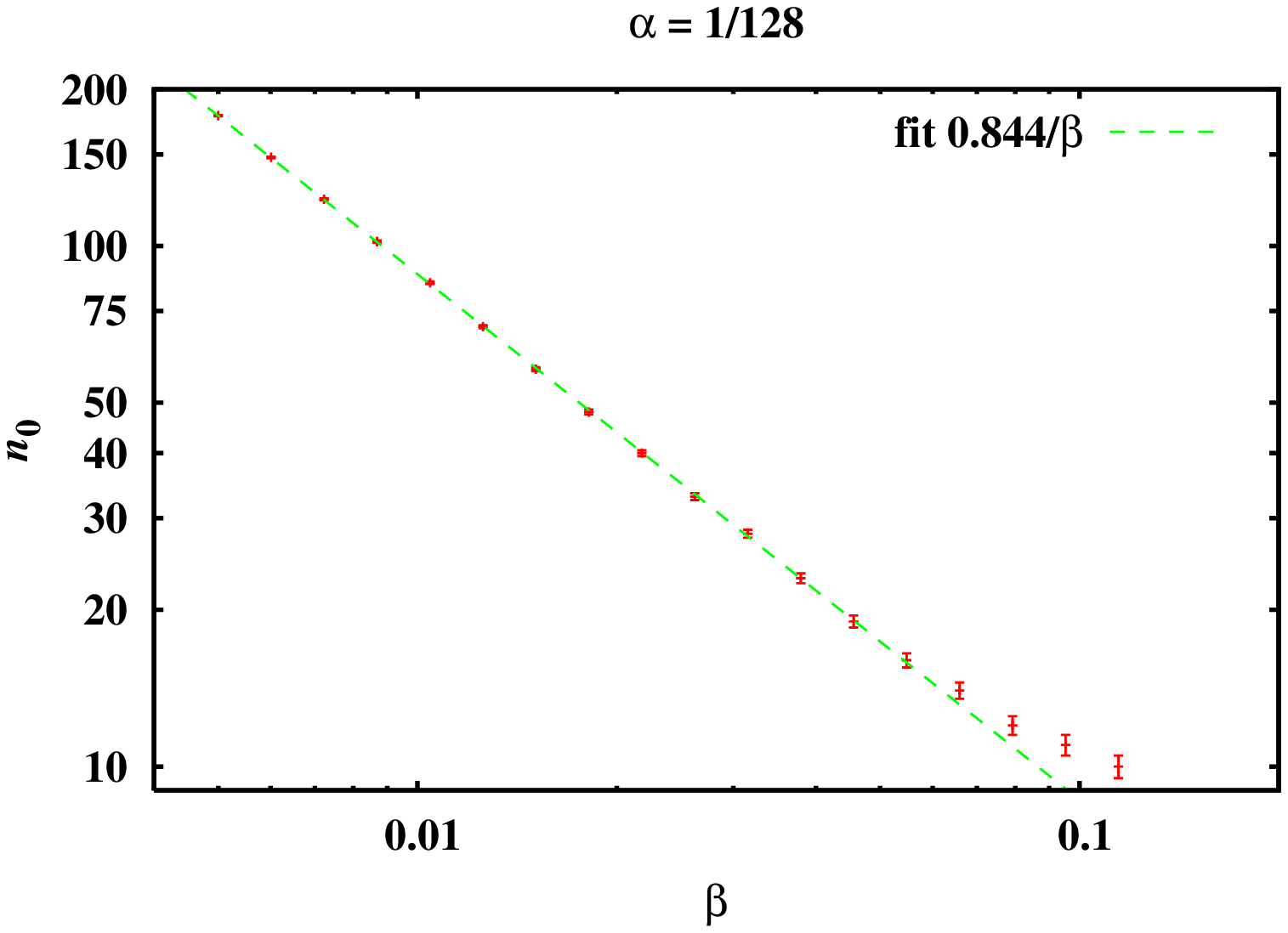}}
\vspace{20pt}}
\mbox{\subfigure
{\includegraphics[width=0.68\linewidth]{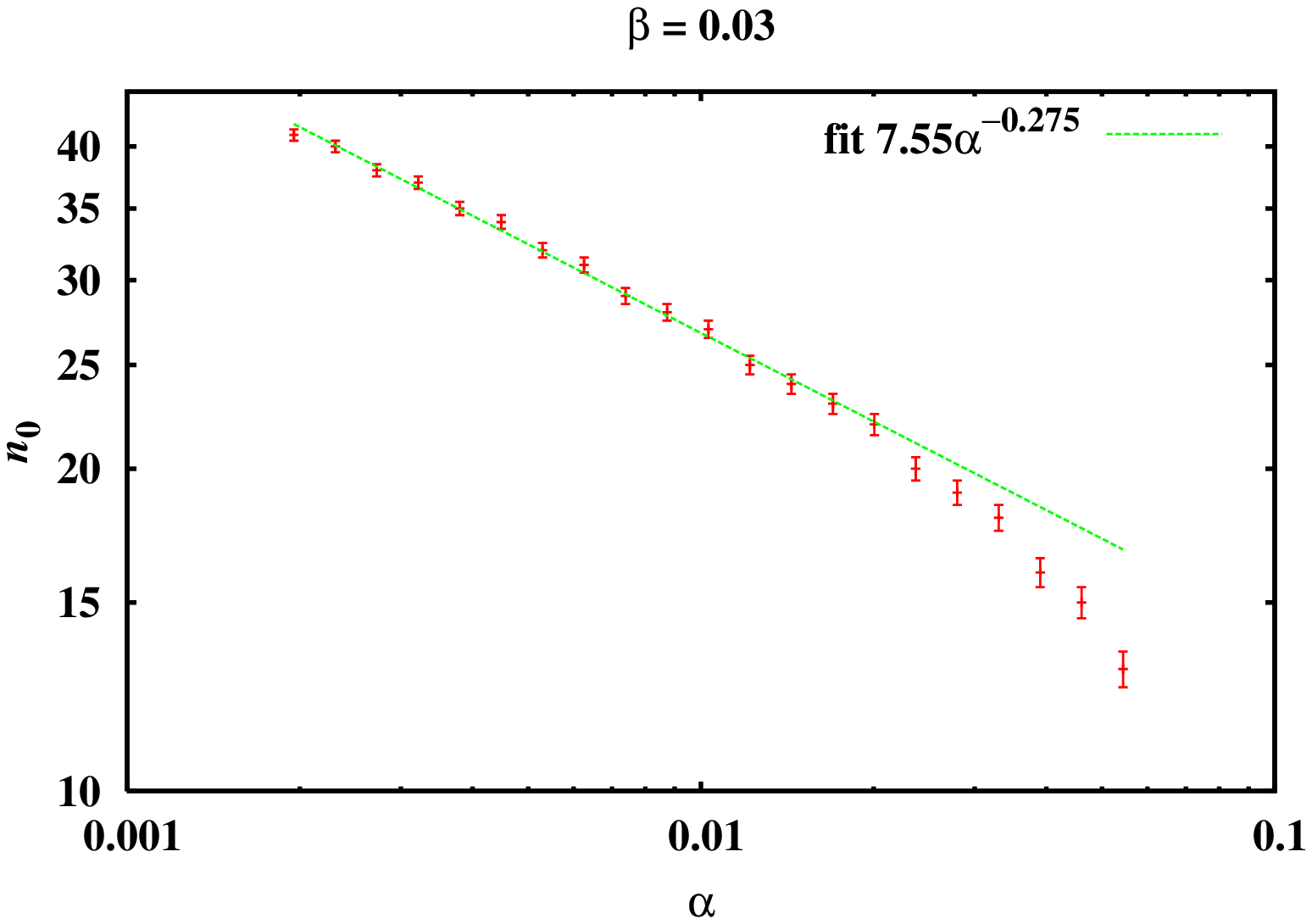}}}
\caption{\protect\small The winding number $n_0$ where the vortex has
minimal vortex tension per unit flux $\mathcal{T}$. \emph{Top panel}: $n_0$
as function of the coupling 
$\beta$ for fixed coupling $\alpha=\frac{1}{128}$. For small $\beta$
($\beta \lesssim 0.05$), the fit shows that $n_0$ scales quite well
proportional to $\frac{1}{\beta}$. Note that the potential is such
that the Chern-Simons vacuum becomes unstable for $\beta \ge
\frac{1}{3}$. \emph{Bottom panel}: $n_0$ as function of the coupling
$\alpha$ for fixed coupling $\beta=0.03$. For small $\alpha$ ($\alpha
\lesssim 0.02$), the fit shows good scaling proportional to
$\alpha^{-\frac{11}{40}}$. The error-bars are simply a reminder that
the function $n_0 \in \mathbb{Z}$.}
\label{fig:nzero}
\end{center}
\end{figure}
\begin{figure}[h!tb ]
\begin{center}
\includegraphics[width=0.7\linewidth]{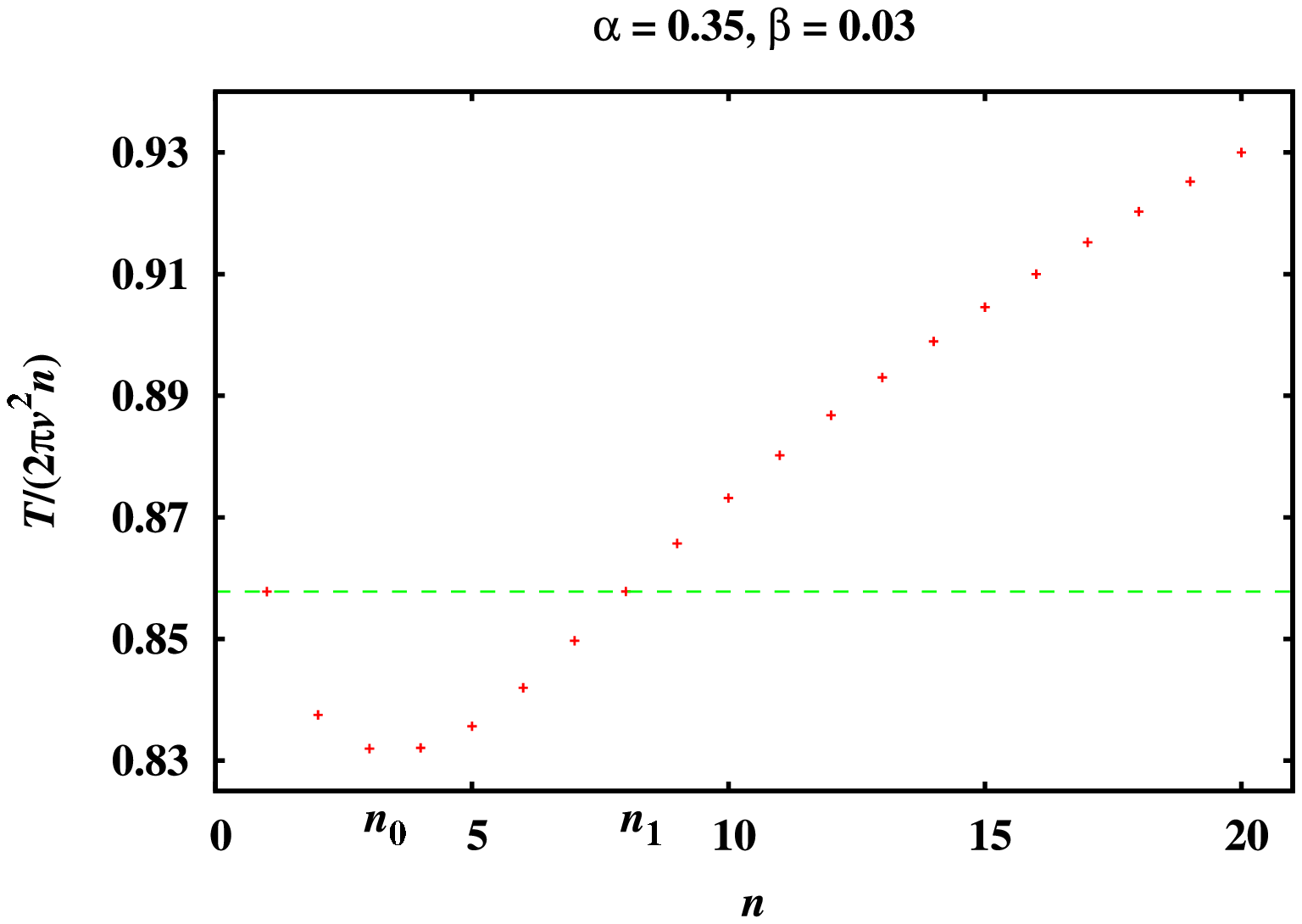}
\end{center}
\caption{{\protect\small Vortex tension divided by $2\pi v^2 n$ as function
    of the winding number $n$, for a vortex with $\alpha = 0.35$ and
    $\beta = 0.03$. The couplings are tuned in such away that $n_0$,
    the winding with minimal vortex tension per unit flux, is
    very small (here $n_0=3$). Thus the optimal
    size, energetically, is quite small, but still bigger than the
    1-vortex. A simple calculation shows that all vortices with $n>4$
    are unstable to decay. }}
\label{piccola}
\end{figure}

In domain $C$ it is trivially shown that all vortices are unstable to
decay into 1-vortices : $T(n) > nT(1), \ \forall n\ge n_1$.

In domain $B$ it is a priori not so easy to see which channels are
allowed and depends on the numerics. In general, the vortices will be
unstable to decay in some channels, but it is not certain that there
cannot be stable vortices here. We can comment on special points which
are instable, that
is, one can easily show that $T(r n_0) > r T(n_0), r\in\mathbb{Z}_+$,
however, these windings might be larger than $n_1$. 

The upshot is to note that for small $\alpha$ and $\beta$, there
will exist a ``fat'' (winding $>1$) vortex with a finite winding
number $n_0$ which is stable and preferred energetically and above a
certain winding number larger than $n_0$ the vortices will decay. This
means that vortices will attract to some certain finite size and
could be detectable in certain kinds of superconductors and
superfluids in $2$ dimensional systems. In other words, the vortices
are attractive until
they reach a critical size and then from that point they will repel
additional fluxes. Hence, it is type I at first and when flux adds up,
it turns into type II, we could denote this behavior a type III
vortex.$\,$\footnote{In non-Abelian non-BPS theories there are more
  possibilities as the forces in general have dependence on the
  internal properties of such systems. Recently, it was shown
  \cite{Auzzi:2007iv} that such a non-Abelian vortex in an
  Extended-Abelian-Higgs theory can have a distance dependent force
  which turns from attractive to repulsive at some distance. }

We explore now an approximate behavior of the function
$n_0(\alpha,\beta)$. In Figure \ref{fig:nzero}, we show the winding
number $n_0$ where the vortex has minimal vortex tension per unit flux (see
Figure \ref{nonzerobetaprima}). In the top-most panel is shown $n_0$ as
function of $\beta$ for fixed $\alpha = 1/128$ and in the bottom-most
panel, $n_0$ as function of $\alpha$ for fixed $\beta = 0.03$. In
both figures, we have made a fit valid for small values of
$\beta,\alpha$, respectively.

Around the point $\left(\frac{1}{128},0.03\right)$ in
$(\alpha,\beta)$-space, we can from the fits guess the following
approximate formula, which is only valid for small couplings (as the
effect is terminated when $\alpha\sim\alpha_{\rm critical}$ or 
$\beta\sim\beta_{\rm critical}$ i.e. $n_0$ becomes equal to one)
\beq n_0 \sim \frac{C}{\alpha^{\frac{11}{40}}\beta}
\ , \eeq
where the constant is $C \sim 0.22$. We have found
$\alpha_{\rm critical}$ and $\beta_{\rm critical}$ to be less than one
but of order $\mathcal{O}(10^{-1})$. It could be interesting to see
how these functions : $\alpha_{\rm critical}(\beta), \beta_{\rm
  critical}(\alpha)$ behave, but would require a better understanding
of the effects coming to play at large couplings.

For phenomenological considerations, it would be interesting to tune
$n_0$ to some small value. As $n_0$ approaches infinitely large
values, the superconductor is effectively of type I and the flux will
break the superconducting phase. We expect more than a single point in
$(\alpha,\beta)$-space to satisfy this condition, actually a line (region)
near the 
critical border : $\alpha(\beta) \lesssim \alpha_{\rm critical}(\beta)$ or
equivalently $\beta(\alpha) \lesssim \beta_{\rm critical}(\alpha)$.
In Figure \ref{piccola} is shown such a configuration where we have
tuned the parameters as : $\alpha=0.35$ and $\beta=0.03$.

We think that this object could be detectable in 2-dimensional
superconductors in the laboratory, at least in principle.

\section{Abelian Maxwell-Chern-Simons System}
\label{secondasezione}

Adding  a Maxwell term to the Chern-Simons Lagrangian, we have a
richer system which will contain both Abrikosov-Nielsen-Olesen (ANO)
vortices and Chern-Simons vortices in respective limits of the
coupling constants \cite{Maxwellchernsimons}. 
First we will make a short review of the model and then construct the
domain wall and then use the latter to study the vortex in the large
flux limit.

The self-dual Maxwell-Chern-Simons Lagrangian
is \cite{Maxwellchernsimons}
\begin{align}
\mathcal{L} &=-\frac{1}{4}F_{\mu\nu}F^{\mu\nu}+\frac{1}{4}%
\kappa\epsilon^{\mu\nu\rho}A_{\mu}F_{\nu\rho} \nonumber\\
&\phantom{=}\ +\left\vert D_{\mu}\phi\right\vert ^{2}+\frac{1}{2}%
\left(\partial_{\mu}N\right)^{2} -U_{\mathrm{BPS}}\left(\left\vert\phi
\right\vert ,N\right) \ , \label{LMCS}
\end{align}
where the BPS potential reads
\begin{equation}
U_{\mathrm{BPS}}\left( \left\vert \phi\right\vert ,N\right) = \frac{1}{2}%
\left[e\left|\phi\right|^2+\kappa N-ev^2\right]^2 +
e^2N^2\left\vert\phi\right\vert^2 \ .
\end{equation}
Note that in order to obtain a self-dual theory, we have to introduce a
neutral real scalar field $N$. The interesting point to note about
this theory is that it reduces to the Abelian-Higgs theory and the
Chern-Simons theory in respective limits of the couplings
$e,\kappa$.

The theory has two degenerate vacua; a symmetric one where $\phi=0$ and
$N=\frac{ev^2}{\kappa}$ and an asymmetric one where $|\phi|=v$ and
$N=0$. Topological solitons exist in the
asymmetric phase and so-called non-topological solitons exist in the
symmetric phase, see e.g. ref. \cite{Maxwellchernsimons}. In order to
see that the system reduces correctly to the two theories, one can
perform the Bogomol'nyi completion of the energy
\begin{align}
E = \int\,d^2x\ \bigg\{&
\frac{1}{2}\left(F_{i0}\pm\partial_iN\right)^2
+\frac{1}{2}\left(F_{12}\pm\left(e|\phi|^2+\kappa
N-ev^2\right)\right)^2 \nonumber\\
&+\left|D_0\phi\mp ie\phi N\right|^2
+\left|D_\pm\phi\right|^2
+\frac{1}{2}\left(\partial_0N\right)^2\bigg\}
\pm ev^2\Phi_B \ ,
\end{align}
from which the self-duality equations are obtained
\begin{align}
D_\pm\phi = 0\ , \label{mcs-Dpm} \\
F_{12} \pm \left(e|\phi|^2-ev^2+\kappa N\right) = 0\ , \label{mcs-self-dual}
\end{align}
where $A_0 = \mp N$ solves two of the four BPS equations. There are
too many degrees of freedom in this 
system and it needs to be accompanied by the Gauss law, which is
simply the variation of the Lagrangian (\ref{LMCS}) with respect to $A_0$
\beq -\partial_iF_{i0} + \kappa F_{12} +ie\left[\phi^*D_0\phi -
\phi (D_0\phi)^*\right] = 0 \ . \label{MCSGaussLaw}\eeq
It is noteworthy to remark that the equation which in the Chern-Simons
case was just algebraic has now become a differential equation.

Back to the two mentioned limits of this theory, it is now easy to see
that the system of BPS equations (\ref{mcs-Dpm}),(\ref{mcs-self-dual})
reduces to that of the Abelian-Higgs model by setting $\kappa=0$, which
in turn allows us 
to set $A_0=N=0$ and the second BPS equation reads
\beq F_{12}\pm e\left(|\phi|^2-v^2\right)=0\ . \eeq
Taking now the other limit
i.e. $\kappa\to\infty$ while holding $\frac{e^2}{\kappa}$ fixed, the
second derivative in the Gauss law (\ref{MCSGaussLaw}) is eliminated
and the Gauss law constraint is the well-known algebraic expression of
the Chern-Simons theory (\ref{gausslaw3}). Henceforth, it is
straightforward to find that the second BPS equation is
\beq F_{12}\pm\frac{2e^3}{\kappa^2}\left(|\phi|^2-v^2\right)|\phi|^2 =
0\ . \eeq

Our immediate aim is to consider the large flux limit of this system.
For the existence of two vacua, it is natural to conjecture that the
compactified domain wall shall be the solution for a large flux
soliton. But what we are particularly interested in, is the matching
of the large-$n$ ANO vortex with the large-$n$ Chern-Simons vortex. We
clearly expect a smooth interpolation between these systems also to be
present in the large flux limit.

Due to the apparent complexity of the system (four degrees of
freedom), from a numerical point of view, we take as a conjecture that
also this system will be well described by a domain wall interpolating
the two vacua: the symmetric and the asymmetric one, being
compactified on a circle. Thus for the topological vortex, which
we consider now, the symmetric vacuum will be enclosed inside the wall
and the asymmetric left outside. Therefore we will study the
Maxwell-Chern-Simons domain wall in the next Subsection.

\subsection{Domain wall}

Considering a dimensional reduction of the system (\ref{LMCS}), we
find the domain wall energy

\begin{align}
T & =\int dx \left[ \left\vert \partial_x\phi \right\vert^2 + \frac{1}{2}%
\left(\partial_xN\right)^2 + U_{\mathrm{BPS}}\left( \left\vert
\phi\right\vert ,N\right)\right]\ , \nonumber\\
&=\int dx \bigg[ \left\vert \partial_x\phi \pm eN\phi\right\vert^2+\frac{1}{2}%
\left(\partial_xN \pm \left(e|\phi|^2+ \kappa N
-ev^2\right) \right)^2  \notag \\
&\phantom{=\int dx \bigg[}
\ \mp \partial_x\left(eN|\phi|^2+\frac{1}{2}\kappa N^2
-ev^2 N \right)\bigg] \ ,
\end{align}
from which the tension of the wall (when it is BPS saturated) can be
read off the total derivative term to be
\begin{equation}
T_{\rm wall} = \frac{e^2v^4}{2\kappa} \ .
\end{equation}
It would be interesting to add flux to the wall, analogously to the
Chern-Simons wall considered previously. We take the energy stemming
from the system (\ref{LMCS}) with the gauge fields turned on. $A_x$
turns out to be zero in the BPS wall and the energy reads
\begin{align}
T & =\int dx \bigg[ \left\vert \partial_x\phi \right\vert^2 + \frac{1}{2}%
\left(\partial_xN\right)^2 + U_{\mathrm{BPS}}\left( \left\vert
\phi\right\vert ,N\right) \nonumber \\
&\phantom{=\int dx \bigg[}
+\frac{1}{2}(\partial_xA_0)^2 +
\frac{1}{2}(\partial_xA_y)^2 + e^2A_0^2|\phi|^2 + e^2A_y^2|\phi|^2
\bigg]\ , \nonumber\\
&=\int dx \bigg[ \left\vert \partial_x\phi \pm eN\phi\right\vert^2+\frac{1}{2}%
\left(\partial_xN \pm \left(e|\phi|^2+ \kappa N
-ev^2\right) \right)^2  \notag \\
&\phantom{=\int dx \bigg[}
\ \mp \partial_x\left(eN|\phi|^2+\frac{1}{2}\kappa N^2
-ev^2 N \right)
+\frac{1}{2}\left(\partial_xA_0 \pm' \partial_xA_y\right)^2 \nonumber
\\
&\phantom{=\int dx \bigg[}
+\left((A_0 \pm' A_y)e|\phi|\right)^2
\mp'\partial_x\left(A_y\partial_xA_0 - \frac{1}{2}\kappa A_y^2\right)
\bigg] \ , \label{mcsbpstension}
\end{align}
where the Gauss law has been used in writing the Bogomol'nyi
completion and the second set of $\pm$s is independent of the
first set and thus is marked with a prime. If we now consider a flux
analogous to the previous case of the Chern-Simons wall,
i.e. $A_y(-\infty) = -f$ and $A_y(+\infty) = 0$ and furthermore
$\partial_xA_y(\pm\infty) = 0$. The wall tension changes according
to
\beq
\label{tensionMCSwall} T_{\rm wall+flux} = \frac{e^2v^4}{2\kappa} + \frac{\kappa f^2}{2} \
. \eeq
Notice that the tension coincides with that of the Chern-Simons wall
(\ref{tensione}).

From the tension (\ref{mcsbpstension}) we can read off the BPS
equations of motion. Firstly, we notice the triviality of the flux
equations which simply imply $A_0 = \mp A_y$. From the scalars, on the
other hand, there is interesting information to gather
\begin{align}
&\Xi = \mp \frac{\kappa}{e}\partial_x \chi \ ,\label{Neq} \\
&\partial_x^2 \chi \pm \kappa\partial_x\chi \label{mcsdweq}
-e^2v^2(e^{2\chi}-1) = 0 \ ,
\end{align}
where $\chi = \log\big(\frac{|\phi|}{v}\big)$ and $\Xi = \kappa
N$. Let us first look
at the limits of the
wall. Taking $\kappa\to\infty$ holding $\frac{e^2}{\kappa}$ fixed, we
readily obtain exactly the Chern-Simons wall
(\ref{basiccswall}). Taking now $\kappa\to 0$ we see the system is
governed by the mass-squared $e^2v^2$ which stems from the Abelian-Higgs
system.

We now are now settled to solve the domain wall (\ref{mcsdweq})
numerically. The scalar field $N$ is readily obtained from (\ref{Neq})
and the magnetic field has to be obtained by the Gauss law
(\ref{MCSGaussLaw}) with $A_0 = A_y$. It is noteworthy to remark that
the vorticity of the uncompactified wall is not fixed a priori, so a
rescaling of the magnetic field is a degree freedom of this
system. However, upon compactification for obtaining the wall vortex, the flux
density is simply $f=\frac{m}{2e}$ as was computed for the Chern-Simons
domain wall (\ref{flux-density}). Hence, the vorticity is given in the
large flux limit and is not a free parameter.

Looking ahead of view, we also know that the BPS equations for the vortex
system should be obeyed for the vortex, which in the large flux limit
looks like a wall, and we could make the crude guess to use the
equation for the magnetic field (\ref{mcs-self-dual}), which is purely
algebraic. We find that the result coincides with the integration of
the Gauss law, up to an
overall factor, but this simply fixes correctly the vorticity such
that (\ref{flux-density}) is obeyed.

\begin{figure}[h!tp]
\centering
\subfigure
{\includegraphics[width=0.48\linewidth]{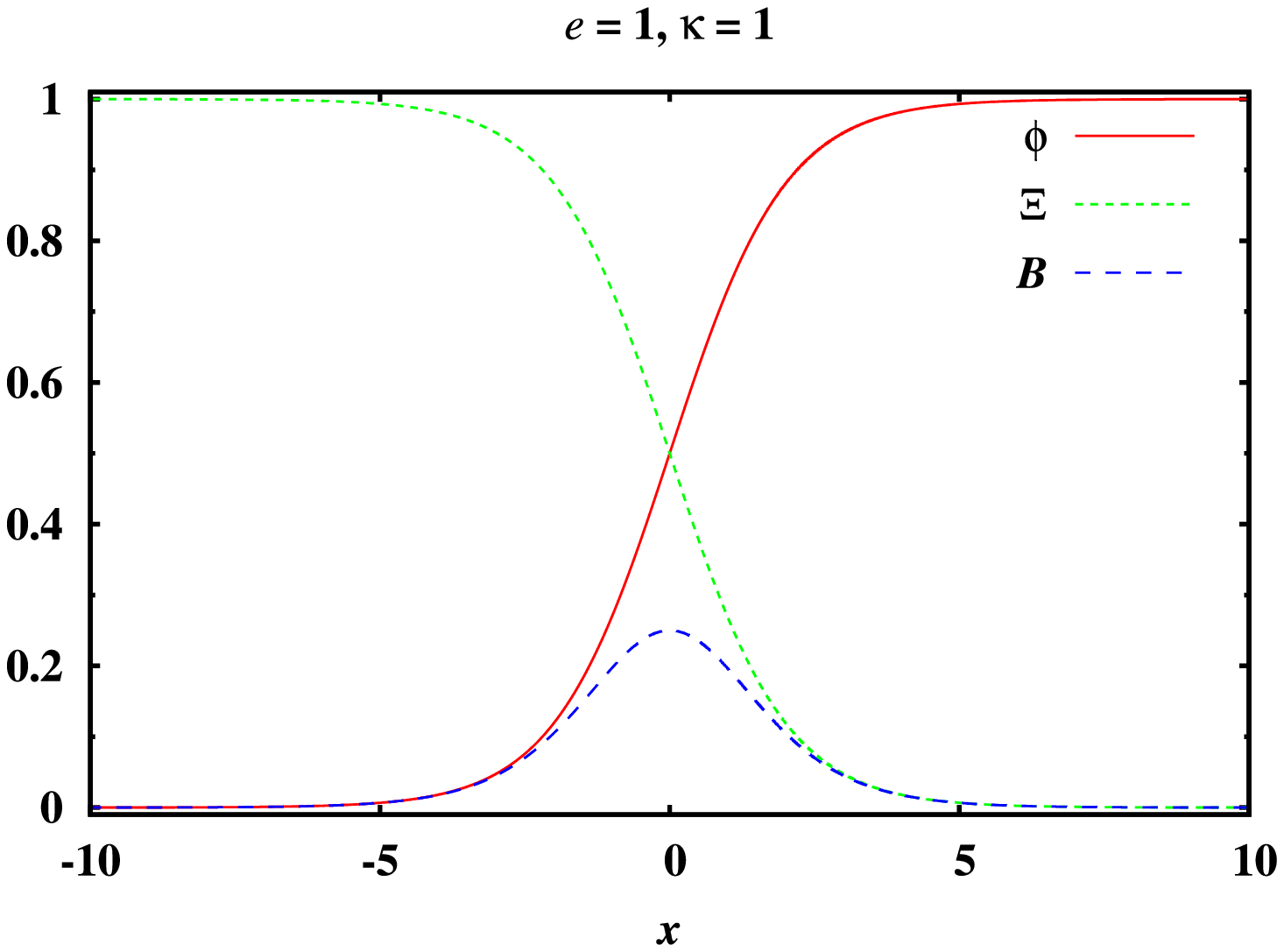}}\quad
\subfigure
{\includegraphics[width=0.48\linewidth]{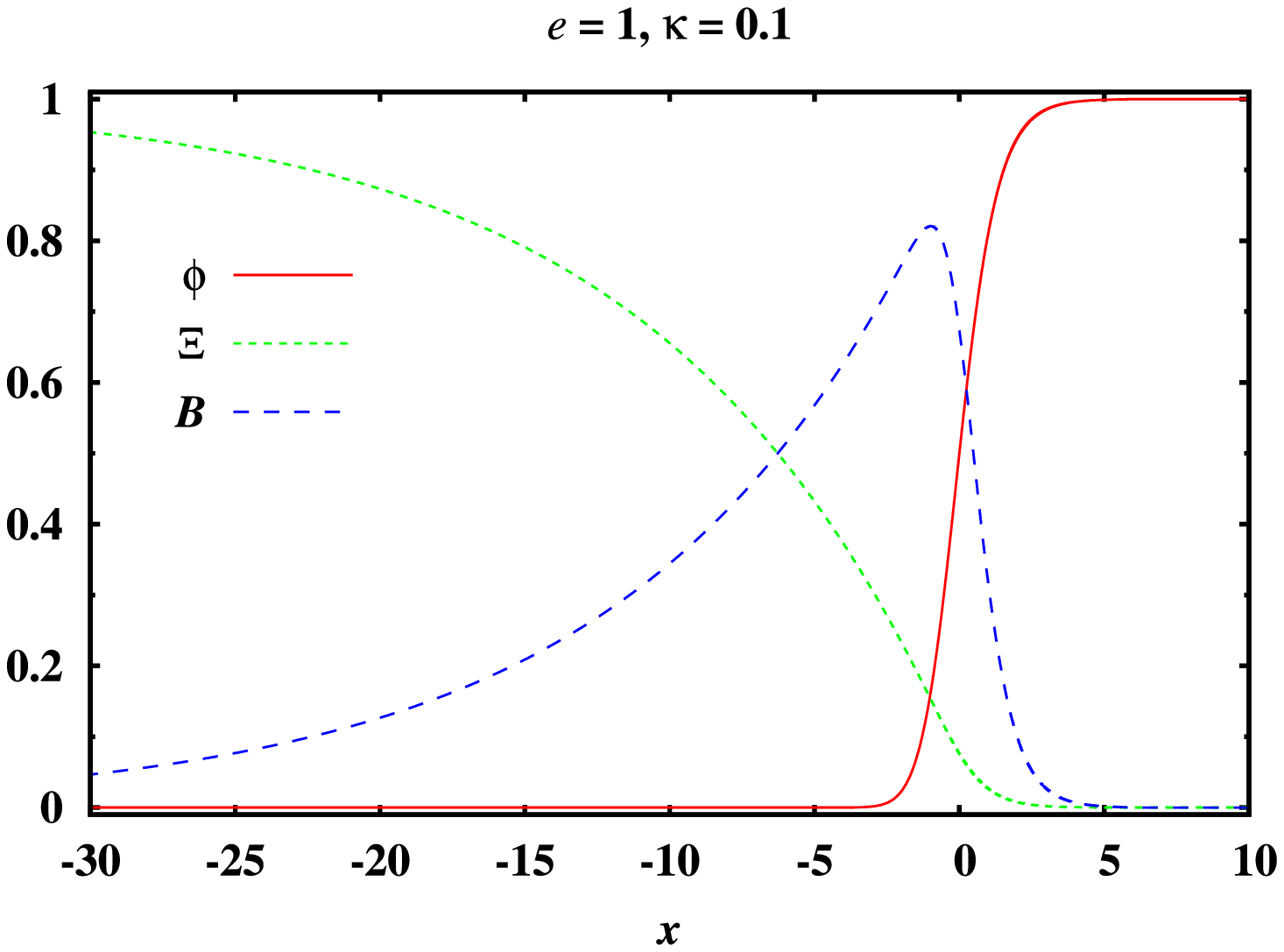}}
\caption{\protect\small The fields of the Maxwell-Chern-Simons domain
  wall with corresponding magnetic field properly normalized. Left
  panel: $e=1,\kappa=1$. Right panel: $e=1,\kappa=0.1$. Notice how the
  magnetic field starts to be pushed ``inside'' the domain wall
  (inside means inside the wall vortex, when one considers a
  compactification of the domain wall on a circle). }
\label{fig:mcswall1}
\end{figure}

\begin{figure}[h!tp]
\centering \subfigure
{\includegraphics[width=0.63\linewidth]{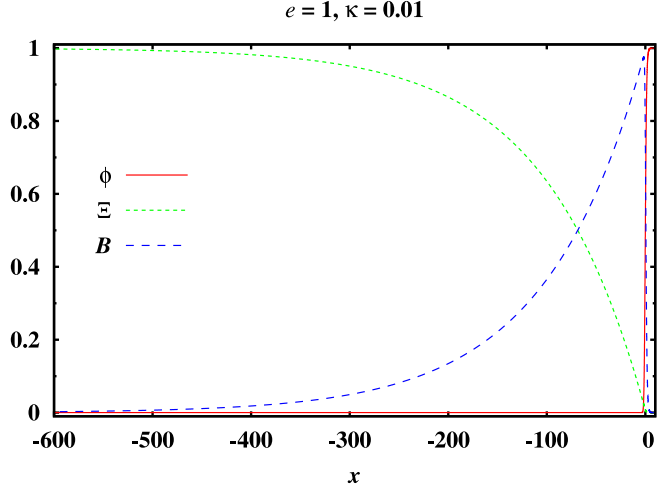}}
\caption{\protect\small The tails of the fields of the
  Maxwell-Chern-Simons domain wall for $e=1,\kappa=0.01$ with
  corresponding magnetic field properly normalized. Solving the BPS
  equation for $N$ from (\ref{mcsbpstension}) with $\phi=0$ we obtain
  the tail (which means far to the left of the wall) of the field
  $N\sim\frac{ev^2}{\kappa}\left(1-e^{\kappa x}\right)$ and thus the
  tail of the magnetic field goes like $B\sim ev^2e^{\kappa
  x}$.}
\label{completo}
\end{figure}

\begin{figure}[h!tp]
\centering
\subfigure
{\includegraphics[width=0.48\linewidth]{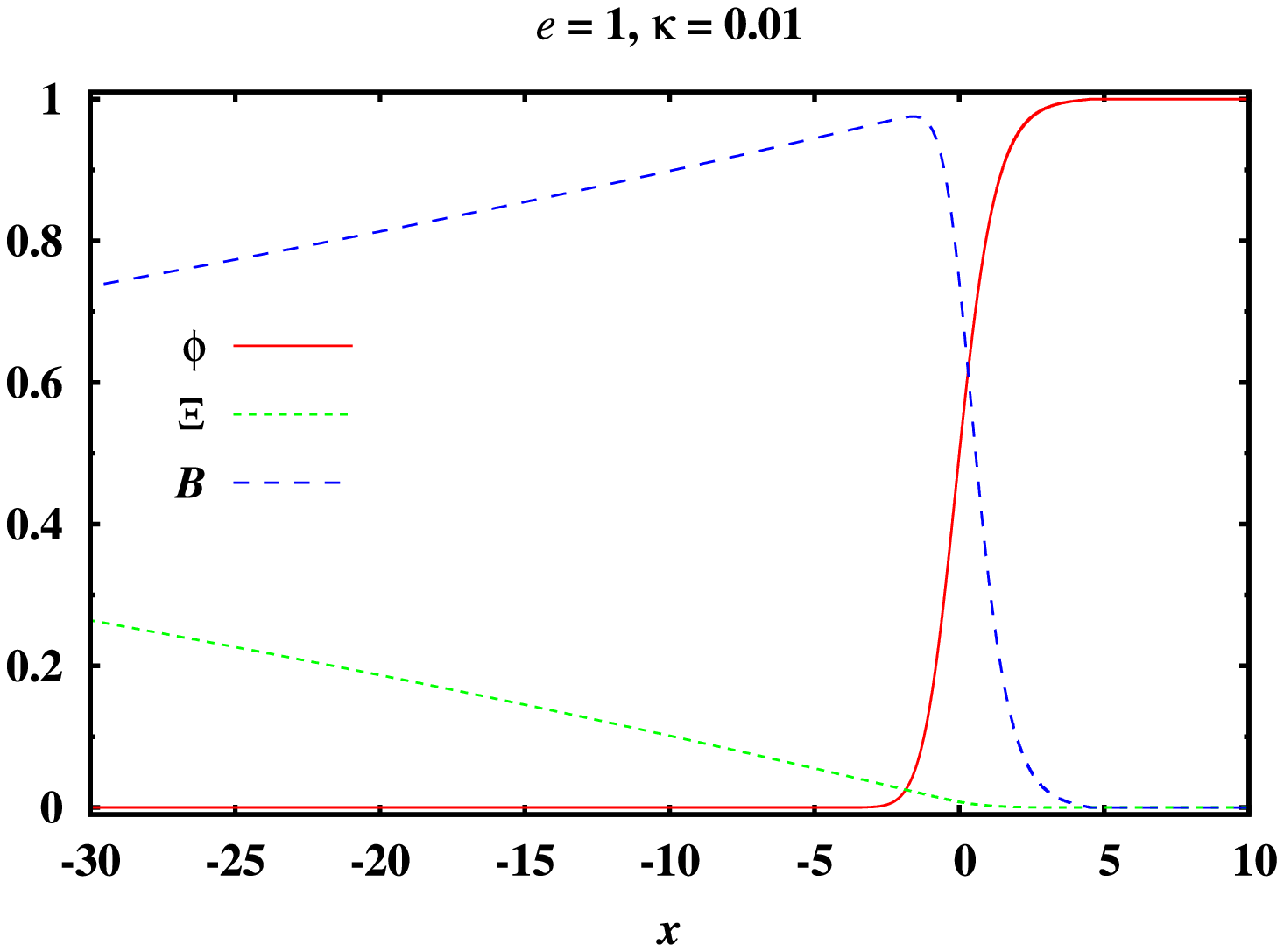}}\quad
\subfigure
{\includegraphics[width=0.48\linewidth]{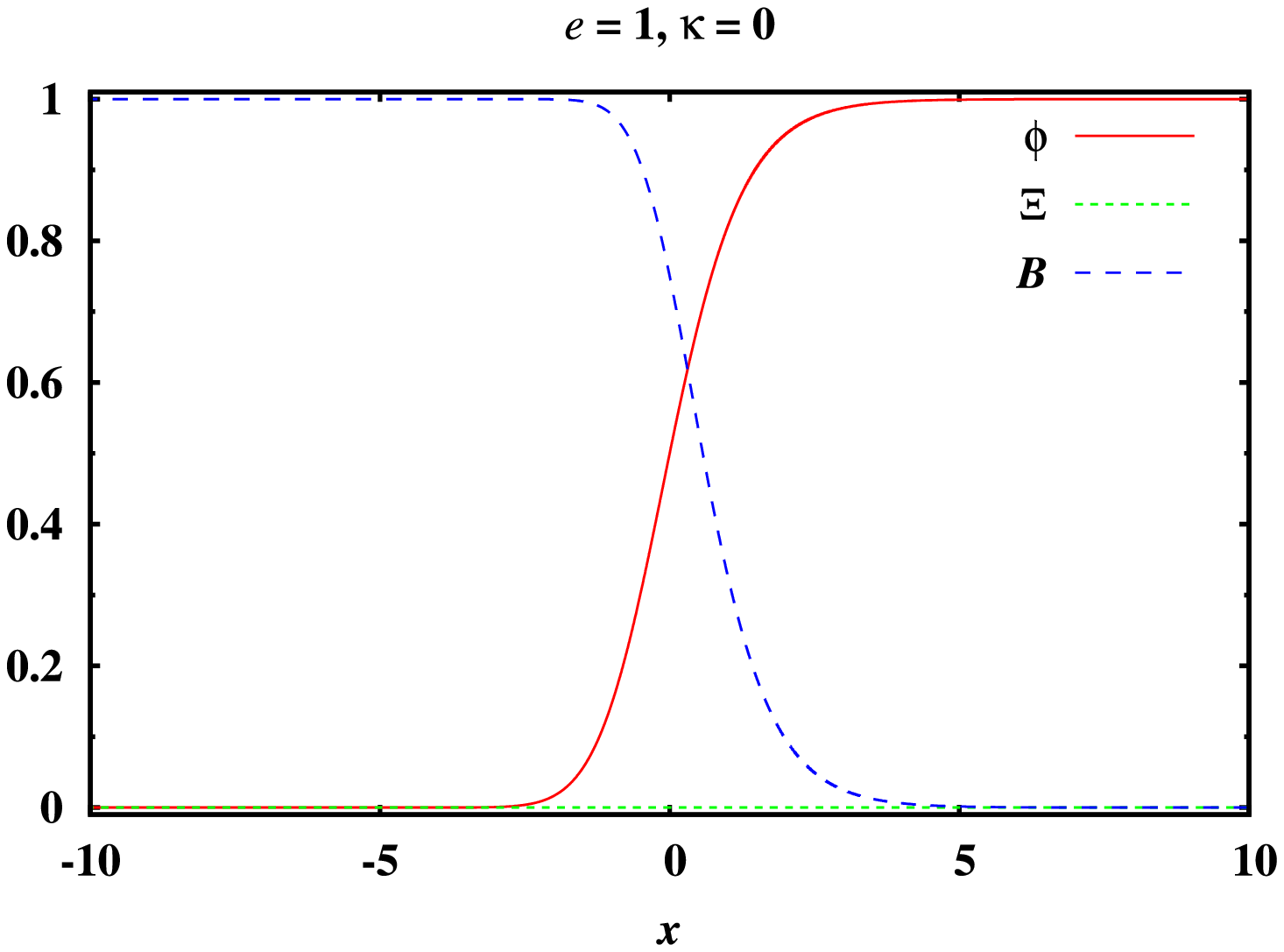}}
\caption{\protect\small The fields of the Maxwell-Chern-Simons domain
  wall with corresponding magnetic field properly normalized. Left
  panel: $e=1,\kappa=0.01$. Right panel: $e=1,\kappa=0$. Notice how
  the field $\Xi$ goes slowly towards its VEV (here
  $\langle \Xi \rangle = 1$) and then for $\kappa\to 0$ the VEV is
  pushed off to infinity and the Coulomb phase can persist inside the
  wall vortex. }
\label{fig:mcswall2}
\end{figure}

The numerical results are shown in Figures \ref{fig:mcswall1},
\ref{completo} and \ref{fig:mcswall2}. We start by entering the
scene with couplings $e=\kappa=1$ and we notice a Chern-Simons-like
behavior of the fields: the magnetic field is trapped in the wall.
Keeping the electric charge fixed and decreasing the Chern-Simons
coupling $\kappa$ we observe that the magnetic field crawls inside
the wall, i.e. to the side of the symmetric phase. This is expected,
as the magnetic flux is captured inside the ANO vortex, and we
expect to see a plateau rising as $\kappa$ goes to zero. Taking a
look at the left hand side graph of Figure \ref{fig:mcswall2}, we
see that some plateau is forming, but the field $\Xi = \kappa N$
wants to get back to its VEV, which is $ev^2$ (here
$\langle\Xi\rangle = 1$). As $\kappa$ is sent to zero the VEV of
$\Xi$ is pushed away to infinity and the ANO wall vortex has
emerged. Note that the energy is infinite, unless the wall is
compactified on a circle.

\subsection{In the large $n$ limit of the vortex:\\ 
interpolation between ANO type and CS type}

Let us now go back to the vortex system
(\ref{mcs-Dpm})-(\ref{mcs-self-dual}). It will prove convenient to
introduce the dimensionless parameter $\eta=\frac{\kappa}{ev}$. This
parameter governs the transition between the pure Maxwell theory
($\eta=0$) and the pure Chern-Simons theory ($\eta \to \infty$). For
all values of $\eta$ there exists a topological vortex. We already
know \cite{Jackiwesteso}, that in the large $n$ limit the Chern-Simons
vortex ($\eta \to \infty$) behaves like a ring of radius $R_V =
\frac{\kappa n}{e^2v^2}$. The ring is the domain wall that separates
the symmetric phase from the Higgs phase. 
For a generic Maxwell-Chern-Simons vortex, we can use the same argument
of \cite{Jackiwesteso} to 
understand the large $n$ limit. As in the pure Chern-Simons case, we
still have a symmetric vacuum and a Higgs vacuum.
We also have a domain wall between the two vacua and, from the
analysis of the previous section, we know that the wall can support a
magnetic flux. We can thus conclude that the large $n$ limit will 
always be a ring-like structure made of the domain wall. A stabilization
computation using formula (\ref{tensionMCSwall}) gives the correct
tension for the vortex.

\begin{figure}[h!tb]
\centering
\includegraphics[width=0.70\linewidth]{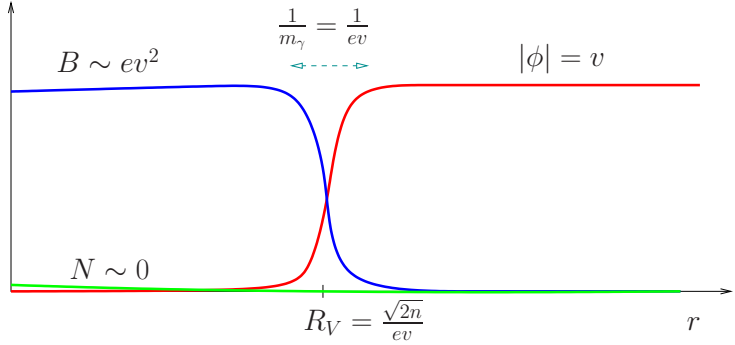}
\caption{\protect\small The profile functions of the
  vortices in the ANO-disc phase. The magnetic field is a
  plateau making up the flux tube. The field $N$ is almost zero in
  this phase. }
\label{fig:anotype}
\end{figure}

For the pure Abrikosov-Nielsen-Olesen vortex, the behavior in the
large $n$ limit 
is completely different \cite{miei}: it becomes a disc of radius $R_V
= \frac{\sqrt{2n}}{ev}$ with the magnetic flux uniformly distributed
inside (Fig.~\ref{fig:anotype}).   
We now want to understand the transition between these two different
regimes. Since we can interpolate smoothly between pure Chern-Simons  
and pure Maxwell theory by changing the parameter $\eta$, it must
certainly be possible to smoothly interpolate between the disc ($R_V
\propto \sqrt{n}$) and the ring ($R_V \propto n$) phases.

\begin{figure}[h!t]
\centering
\includegraphics[width=0.90\linewidth]{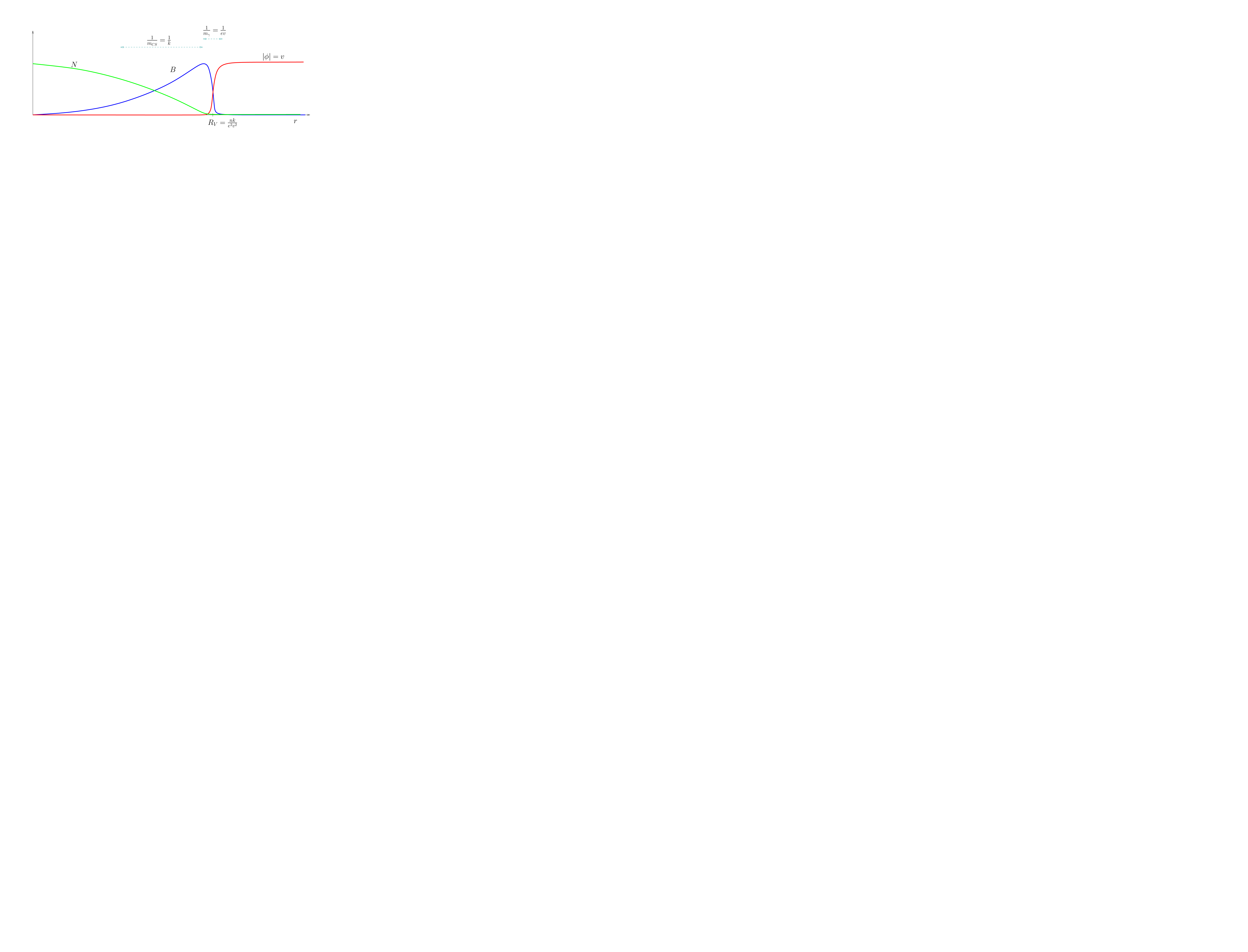}
\caption{\protect\small A sketch of the profile functions of the
  vortices in the CS-ring phase. The magnetic field will have an
  exponential from the ring, where the size of the ring is the inverse
  photon mass $1/\kappa$. The approximate profile of the magnetic
  field is $B = ev^2 e^{\kappa (r-R_V)}$ while 
  $N=ev^2\left(1-e^{\kappa (r-R_V)}\right)/\kappa$.}
\label{fig:cstype}
\end{figure}

As already mentioned, it is quite difficult to attack the problem of
the large $n$ MCS vortex numerically. To (re)construct the phase
diagram ($\eta,n$) we have to use some intuition and extrapolate from
the previous results of the domain wall with magnetic flux. 
In the very large $n$ limit, the vortex always becomes a domain
wall-ring. The discussion (\ref{tensione})-(\ref{flux-density}) done
for the pure Chern-Simons vortex, follows unchanged for the MCS wall
of (\ref{tensionMCSwall}) (see Figure \ref{fig:cstype}). 

The interesting thing happens for sufficiently small $\eta$, when the
theory is almost Maxwell and the Chern-Simons photon mass is very small
compared to the other scales of the theory.  We want to consider what
happens to the vortex in this region, small $\eta$ and progressively
larger $n$.

In the first stage, the vortex behaves like an ANO-type vortex (see
Figure \ref{fig:anotype}). The radius grows like $R_V =
\frac{\sqrt{2n}}{ev}$, such that the area is linear in $n$ and the
magnetic flux is uniform inside the disc. At the edge of the disc
there is a transition between the internal Coulomb phase and the
external Higgs vacuum. The thickness of this transition is of order of
the inverse Higgs photon mass $1/(ev)$. The real scalar field $N$
essentially remains zero at this first stage.  

When the radius of the vortex becomes sufficiently large, reaching the
inverse of the mass of the Chern-Simons photon, $1/\kappa$, the $B$ and $N$
profiles start to develop exponential tails due to the topological
mass. We are entering a transition region between the ANO-type and the 
CS-type behavior.

\begin{figure}[h!]
\centering
\includegraphics[width=0.70\linewidth]{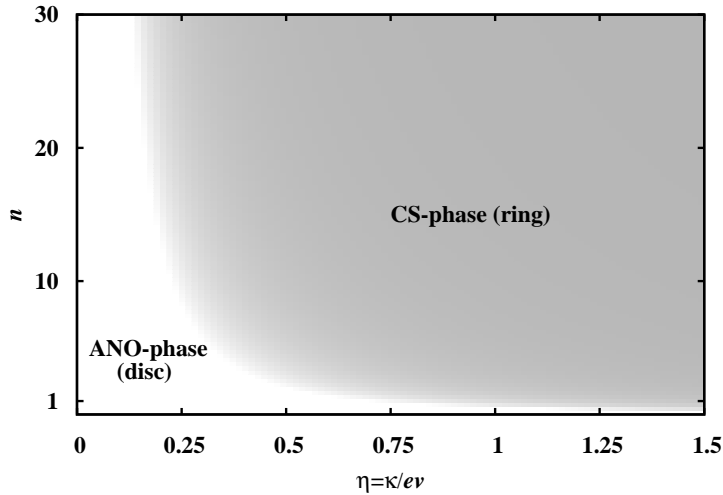}
\caption{\protect\small A phase diagram for the MCS vortex. The
  ANO-type phase and the CS-type phase can be separated by the curves
  $n=1/(2\eta^2)$ and $n=1/\eta^2$. The transition is a smooth
  transition as illustrated by the tone. } 
\label{fig:phases}
\end{figure}

Thus, we can find the ANO-like vortex ``upper bound'' by comparing the
radius of the ANO vortex $R = \frac{\sqrt{2n}}{ev}$ with the inverse
of the CS photon mass, which is $1/\kappa$. The CS-like ``lower
bound'', however, can be found by comparing the radius of the CS
vortex $R = \frac{\kappa n}{e^2v^2}$ with  $1/\kappa$ 
\beq n_{\rm ANO\  upper} = \frac{1}{2 \eta^2} \ , \qquad 
n_{\rm CS\ lower} = \frac{1}{\eta^2} \ . \eeq 
The phase diagram is shown in Figure \ref{fig:phases}.

\section{Conclusions and Further Developments \label{conclusione}}

Firstly, we have shown explicitly, using numerical tools, that in the large
flux limit, Chern-Simons vortices approach closed domain walls.
Next, we have considered the behavior of Chern-Simons vortices with a
general non-BPS potential. Interesting potentials can provide
stabilization for vortices with an arbitrary number, which we denote
$n_0$, of magnetic fluxes. This phenomenon is new from the point of
view of the ordinary ANO vortex, where only type I and type II
behaviors are possible. We could tentatively denote this behavior
a type III vortex. The interesting point is that, in principle, it
could be detectable, as it resembles the physics of a type II
superconductor,
but with the flux tubes being physically bigger. Furthermore, it could
be a nice playing ground for studying vortex interactions
experimentally. We should emphasize once more, that the configurations
need 2 spatial dimensions and most likely complicates the experiment
and thus it could be a challenge for the nanotechnological scene.

Secondly, we have studied the Maxwell-Chern-Simons domain wall, which
we conjecture to resemble the vortex in the large flux limit. With
this at hand, we have shown a smooth interpolation between the
Chern-Simons vortex and the ANO vortex, in the large magnetic flux limit.

There are some interesting lines of future research and
generalizations of these results. One is to consider the non-Abelian
generalization of Chern-Simons solitons. Recently \cite{non-abelian},
it has been shown that a non-Abelian generalization of the
Chern-Simons-Higgs system gives the possibility of internal degrees
of freedom living on the vortex. It would be interesting to explore
also in this case the relation between the domain wall and the
vortices with large flux. It would also be interesting to consider the
addition of a Yang-Mills term and hence study the continuous
interpolation between the two models, pure Chern-Simons
\cite{non-abelian} and pure Yang-Mills \cite{YM}. Another interesting
line of research is the string-theoretical realization of Chern-Simons
solitons \cite{Lee:1999ze}. Again it would be interesting to
understand the large $n$ behavior in this setup. Also interesting are
the non-commutative geometry realizations of vortices, and in
particular the ``puffed'' solution of ref. \cite{jarah}.

\section*{Acknowledgments}

S.B.G. would like to thank Minoru Eto, Kenichi Konishi and Walter
Vinci for fruitful discussions. The work of S.B. is supported by DOE grant DOE/DE-FG02-94ER40823.

\end{document}